\documentclass[aps]{revtex4}

\usepackage{graphicx}

\begin{document}

\title{Random matrix study for a three-terminal chaotic device}


\author{A. M. Mart\'inez-Arg\"uello}
\affiliation{Departamento de F\'{\i}sica, Universidad Aut\'onoma
Metropolitana-Iztapalapa, 09340 M\'exico Distrito Federal, Mexico}

\author{E. Casta\~no}
\affiliation{Departamento de F\'{\i}sica, Universidad Aut\'onoma
Metropolitana-Iztapalapa, 09340 M\'exico Distrito Federal, Mexico}

\author{M. Mart\'inez-Mares}
\affiliation{Departamento de F\'{\i}sica, Universidad Aut\'onoma
Metropolitana-Iztapalapa, 09340 M\'exico Distrito Federal, Mexico}

\begin{abstract}
We perform a study based on a random-matrix theory simulation for a
three-terminal device, consisting of chaotic cavities on each terminal. We
analyze the voltage drop along one wire with two chaotic mesoscopic cavities,
connected by a perfect conductor, or waveguide, with one open mode. This is done
by means of a probe, which also consists of a chaotic cavity that measure the
voltage in different configurations. Our results show significant differences
with respect to the disordered case, previously considered in the literature.
\end{abstract}

\maketitle

\section{Introduction}

In the last thirty years there have been much theoretical and experimental
work concerning electronic transport through multiterminal devices (see
Refs.~\cite{Datta,Goodnick} there in). Nowadays, the interest to study the
electronic transport properties on these devices has been
renewed~\cite{Picciotto,Chang,Geisel,Haan,Gao,Aita}, due to the fact that they
are very useful in experimental measurements in several
configurations~\cite{ButtikerPRA,ButtikerPRL,ButtikerIBM}. 

The earlier experiments were done with normal metal conductors, whose random
distribution of impurities in their microscopic structure, give rise to
interference that is reflected in the relevant physical observables, like 
resistance or voltage measurements. Moreover, those quantities show sample
to sample fluctuations~\cite{GodoyEPL,Godoy,GMM1}. More recently, the interest
on these systems has resurged due to recent advances in technology, that allow
to access to clean devices, where the typical size is smaller than the elastic
mean free path. Therefore, the electrons propagate ballistically and scattering
is produced only by the device boundaries, which have important consequences in
the electronic transport through the device~\cite{Haan,Song1,Song2}. For
instance, when the geometry is such that the classical dynamics in the system is
chaotic, the transport properties fluctuates
too~\cite{MelloLesHouches,BeenakkerRMP,Alhassid}. What is very important is to
know how are the fluctuations with respect to the disordered case.

In this work, by numerical simulation, we analyze the statistical
distribution of the voltage drop along a chaotic wire, which consists of two
chaotic mesoscopic cavities connected by a perfect conductor with one open mode.
The probe is a chaotic cavity that measure the voltage in different
configurations. The presence and absence of time reversal invariance are
considered. We compare our results with the ones obtained in an equivalent three
terminal device but with disordered, instead of chaotic, wires, previously
studied in the literature, where the distribution of the voltage drop was
determined in the presence of time reversal invariance
only~\cite{GodoyEPL,Godoy}. There, a remarkable difference in the distribution
of the voltage drop between the ballistic regime and the strong disordered
limit, has been found. The position of the probe has a stronger effect than in
the disordered case.

First, we summarize the scattering formalism for the voltage drop in a three
terminal device, proposed by B\"uttiker~\cite{ButtikerPRL,ButtikerIBM}. Then, we
construct the scattering matrix for our system, in terms of the scattering
matrices of the individual cavities, as well as of the scattering matrix
associated to the junction, for which we assume the simplest model introduced by
B\"uttiker, that couple the probe symmetrically to the horizontal
wire~\cite{ButtikerPRA}. For the statistical analysis, we make an ensemble of
systems by assuming that the scattering matrix of each cavity is chosen from a
Circular Ensemble, Orthogonal or Unitary, depending on the presence or absence
of time reversal invariance. We present our conclusions at the end.


\section{Electronic transport in a three-terminal system}

In the formulation of Landauer-B\"uttiker, the electronic transport is reduced
to a scattering problem. In a single mode multiprobe devices, the current $I_i$
in a lead $i$ can be written into two components, one being the reflection to
the same lead and the transmission from the others leads to the lead $i$. That
is, $I_i$ is given in terms of the reflection and transmission coefficients,
according to~\cite{ButtikerIBM}
\begin{equation}
I_{i} = \frac{e}{h} 
\left[(1-R_{ii}) \mu_{i} - \sum_{j \neq i} T_{ij} \mu_{j}
\right] , 
\end{equation}
where $\mu_{j}$ is the chemical potential in lead $j$, $R_{ii}$ is the
reflection coefficient to the lead $i$, and $T_{ij}$ represent the transmission
from lead $j$ to lead $i$. These coefficients are given by the scattering
matrix $S$ as $R_{ii}=|S_{ii}|^2$ and $T_{ij}=|S_{ij}|^2$.

\begin{figure}
\includegraphics[width=9.0cm]{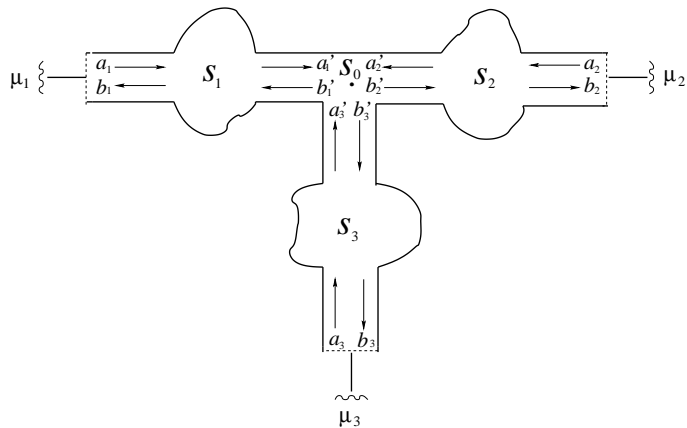}
\caption{A scattering system consisting of three one-dimensional wires
converging to a junction. The thin lines represent perfect conductors
that connect the wave guides to the chemical reservoirs. The amplitude of the
incoming (outgoing) wave in wire $i$ is denoted by $a_{i}$ ($b_{i}$), while
$a'_{i}$ ($b'_{i}$) denotes the amplitude at the junction. Each wire is
described by a $2 \times 2$ scattering matrix $S_{j}$ and the junction by a $3
\times 3$ matrix, $S_{0}$.}
\label{fig:figura1}
\end{figure}

The \emph{voltage} along a horizontal wire, connected via perfect leads
to two reservoirs of fixed chemical potentials, $\mu_{1}$ and $\mu_{2}$, can be 
measured in a three terminal device, where the third wire is in a voltage
measurement configuration (see Fig.~\ref{fig:figura1}); that is, the chemical
potential $\mu_3$ is such that the current through it is equal to zero, $I_3=0$.
In such a case~\cite{ButtikerIBM},
\begin{equation}
\label{eq:chemical}
\mu_{3} = \frac{1}{2} (\mu_{1} + \mu_{2}) + \frac{1}{2} (\mu_{1} - \mu_{2}) f ,
\end{equation}
where $f$ is given by 
\begin{equation}
\label{eq:f}
f = \frac{T_{31} -T_{32}}{T_{31} + T_{32}}.
\end{equation}

Equation~(\ref{eq:chemical}) shows that the chemical potential $\mu_3$ has an
averaged part, that comes from the effect of the resorvoirs $\mu_1$ and $\mu_2$
only. The second part gives the deviation from the averaged part, and depends on
the intrinsic nature of the conductors through the quantity $f$, that contains
all the relevant information about the multiple scattering in the device. If the
conductors are disordered or chaotic, $f$ fluctuates between -1 and 1, since
$\mu_3$ can not reach the values $\mu_1$ nor $\mu_2$ due to the contact
resistance~\cite{ButtikerIBM}. The disordered three terminal device was studied
in Refs.~\cite{GodoyEPL,Godoy}.

In what follows we will consider a three terminal device where the conductors
are chaotic. Since $f$ depends on the scattering matrix $S$ of the whole system,
we construct $S$ in terms of the scattering matrices of each conductor and the
scattering matrix of the splitter, that we will assume to be known. For the
splitter we assume a simple model, while the scattering matrices of the chaotic
conductors are chosen from an ensemble of random matrices that satisfy certain
symmetry requirements. 

\section{The $S$ matrix for a three terminal device}

Let us consider the three terminal system shown in Fig.~\ref{fig:figura1}.
The system is described by the scattering matrix $S$ which relates the incoming
plane waves amplitudes on each terminal, $a_{1}$, $a_{2}$, and $a_{3}$, to the 
outgoing ones, $b_{1}$, $b_{2}$, and $b_{3}$, by
\begin{equation}
\label{eq:S}
\left( \begin{array}{c}
b_{1} \\ b_{2} \\ b_{3}
\end{array} \right)
= S \left( \begin{array}{c}
a_{1} \\ a_{2} \\ a_{3}
\end{array} 
\right),
\end{equation}
where we assume that $S$ contains all the information that from the system we
can obtain. Of course, $S$ depends on the scattering process inside the system,
due to scattering elements. 

Let assume that the splitter is represented by the scattering matrix $S_0$, that
couples the three terminals; therefore, the amplitudes at the junction are
related as 
\begin{equation}
\label{eq:Ssplitter}
\left( \begin{array}{c}
b'_{1} \\
b'_{2} \\
b'_{3}
\end{array} \right)
  = S_{0}
\left( \begin{array}{c}
a'_{1} \\
a'_{2} \\
a'_{3}
\end{array} \right). 
\end{equation}
If the conductors on each terminal are represented by the scattering matrices
$S_j$ ($j=1,2,3$), the amplitudes are related as follows:
\begin{equation}
\label{eq:Scavities}
\left(
\begin{array}{c} 
b_{1} \\ a'_{1} 
\end{array}
\right) = S_1
\left(
\begin{array}{c}
a_{1} \\ b'_{1} 
\end{array}
\right) , \quad
\left(
\begin{array}{c} 
a'_{2} \\ b_{2} 
\end{array}
\right) = S_2
\left(
\begin{array}{c}
b'_{2} \\ a_{2} 
\end{array}
\right) , \quad
\left(
\begin{array}{c} 
a'_{3} \\ b_3 
\end{array}
\right) = S_3
\left(
\begin{array}{c}
b'_{3} \\ a_3 
\end{array}
\right) ,
\end{equation}
where each matrix $S_j$ is a $2\times 2$ matrix with the general structure 
\begin{equation}
\label{eq:Sj}
S_{j} =
\left( \begin{array}{cc}
r_{j} & t'_{j} \\
t_{j} & r'_{j} \\
\end{array} \right),
\end{equation}
with $r_{j}$, $t_{j}$ are the reflection and transmission amplitudes when the
incidence is from the left (or below for $j=3$) of the $j$th conductor, and
$r'_{j}$, $t'_{j}$ when the incidence is from the other side. Flux conservation
implies that $S_j$ is unitary, 
\begin{equation}
\label{eq:properties}
S_j S_j^{\dagger} = I_2 ,
\end{equation}
where $I_2$ stands for the $2\times 2$ identity matrix. Equation
(\ref{eq:properties}) is the only requirement in absence of any symmetry, while
in the presence of \emph{time reversal invariance}, $S_j$ is a unitary and
\emph{symmetric}, 
\begin{equation}
\label{eq:properties2}
S_j = S_j^{T} ,
\end{equation}
where $T$ stands for the transposed.

Through Eqs.~(\ref{eq:Ssplitter}), (\ref{eq:Sj}) and (\ref{eq:Scavities}) we
arrive to the scattering matrix $S$ that describes the full system, which is
given by 
\begin{equation}
\label{eq:Sfull}
S = S_{PP} + S_{PQ} S_{0} \frac{1}{1 - S_{QQ} S_{0}} S_{QP},
\end{equation}
where we have defined 
\begin{equation}
\label{eq:PQ}
S_{PP} = \left(
\begin{array}{ccc}
r_{1} & 0 & 0 \\
0 & r'_{2} & 0 \\ 
0 & 0 & r'_{3}
\end{array}
\right), 
\qquad
S_{PQ} = \left(
\begin{array}{ccc}
t'_{1} & 0 & 0 \\
0 & t_{2} & 0 \\ 
0 & 0 & t_{3}
\end{array}
\right),
\qquad
S_{QP} = \left(
\begin{array}{ccc}
t_{1} & 0 & 0 \\
0 & t'_{2} & 0 \\ 
0 & 0 & t'_{3}
\end{array}
\right),
\qquad
S_{QQ} = \left(
\begin{array}{ccc}
r'_{1} & 0 & 0 \\
0 & r_{2} & 0 \\ 
0 & 0 & r_{3}
\end{array}
\right).
\end{equation}
Equation (\ref{eq:Sfull}) has a nice interpretation: the first term on the right
hand side, $S_{PP}$, represents the reflected parts of the waves that reach the
conductors, while the second term comes from the multiple scattering in the
system. Here, $S_{QP}$ gives the transmission from outside to inside, $S_{PQ}$
gives the transmission from inside to outside of the system, and $S_{QQ}$
represents the internal reflections. 

Notice that $S$ is also a unitary matrix once we ensure that $S_0$ is chosen as
a unitary matrix too, and the symmetry conditions are fixed by the symmetry
properties of the $S_j$'s. Although our result is general, in what follows we
adopt a simple model for $S_0$ and we choose $S_j$ from an ensemble of
scattering matrices that simulates chaotic cavities. 


\subsection{A simple model for the splitter}

A simple model for the $S$-matrix of the splitter, real and symmtric, that
couples the probe \emph{symmetrically}, was proposed by
B\"uttiker~\cite{ButtikerPRA}, namely
\begin{equation}
\label{eq:Smodel}
S_{0} =
\left( \begin{array}{ccc}
a & b & \sqrt{\varepsilon} \\
b & a & \sqrt{\varepsilon} \\
\sqrt{\varepsilon} & \sqrt{\varepsilon} & -(a + b)
\end{array} \right),
\end{equation}
where $\varepsilon$ is a real parameters with $0\le\varepsilon\le 1/2$, which
gives the coupling strength, and 
\begin{equation}
a = -\frac{1}{2}\left(1 - \sqrt{1 - 2\varepsilon}\right), \quad
b = +\frac{1}{2}\left(1 + \sqrt{1 - 2\varepsilon}\right).
\end{equation}
When the coupling vanishes ($\varepsilon\rightarrow 0$), $a\rightarrow 0$ and
$b\rightarrow 1$ which means that the probe is decoupled and there is
complete transmission between the terminals 1 and 2. On the contrary, when the
probe is perfectly coupled ($\varepsilon=1/2$), $a=-1/2$ and $b=1/2$, nothing
is reflected to the probe.

\section{The voltage measurement}

We assume that the conductors in our device are in fact chaotic cavities, such
that the voltage measurement, and any other transport properties, shows sample
to sample fluctuations, although macroscopically seems to be identical; this is
due to the difficulty of control of the shape of the cavity microscopically. Of
course, the fluctuations also arise with respect to external parameters like
the chemical potentials and magnetic field. Therefore, we require to make a
statistical study for the voltage measurement. We do this for two kind of
ensembles for the $S_j$ matrices: in presence and absence of time reversal
symmetry. In the Dyson's scheme these correspond to the orthogonal and unitary
cases, labeled by $\beta=1$ and $\beta=2$, respectively~\cite{Dyson1962}. 

\subsection{Presence of time reversal invariance}

In the $\beta=1$ symmetry, an $S_{j}$ matrix can be parameterized in a 
``polar representation'' as~\cite{MelloLesHouches}
\begin{equation}
\label{eq:Ensemble}
S_{j} = \left(
\begin{array}{cc}
e^{i\phi_{j}} & 0 \\
0 & e^{i\psi_{j}}
\end{array}
\right) \left(
\begin{array}{cc}
-\sqrt{1-\tau_{j}} & \sqrt{\tau_{j}} \\
\sqrt{\tau_{j}} & \sqrt{1-\tau_{j}}
\end{array}
\right) \left(
\begin{array}{cc}
e^{i\phi_{j}} & 0 \\
0 & e^{i\psi_{j}}
\end{array}
\right),
\end{equation}
where $\phi_j$ and $\psi_j$ are random numbers, uniformly distributed in the
interval $[0, 2\pi]$, and $\tau_{j}$ is randomly distributed in $[0,1]$. The
probability distribution for $S_j$ is~\cite{MelloLesHouches}
\begin{equation}
dP_1(S_j) = \frac{d\tau_{j}}{2\sqrt{\tau_{j}}} 
 \frac{d\phi_{j}}{2\pi} 
\frac{d\psi_{j}}{2\pi},
\end{equation}
which defines the \emph{Circular Orthogonal Ensemble}, which can be generated
numerically. Once this is done, we substitute the elements of the $S_j$
matrices in the expressions given in Eq.~(\ref{eq:PQ}), and then in
Eq.~(\ref{eq:Sfull}) from which we obtain the transmission coefficients
$T_{31}$ and $T_{32}$, needed to determine $f$ through Eq.~(\ref{eq:f}).

\begin{figure}
\includegraphics[width=10.0cm]{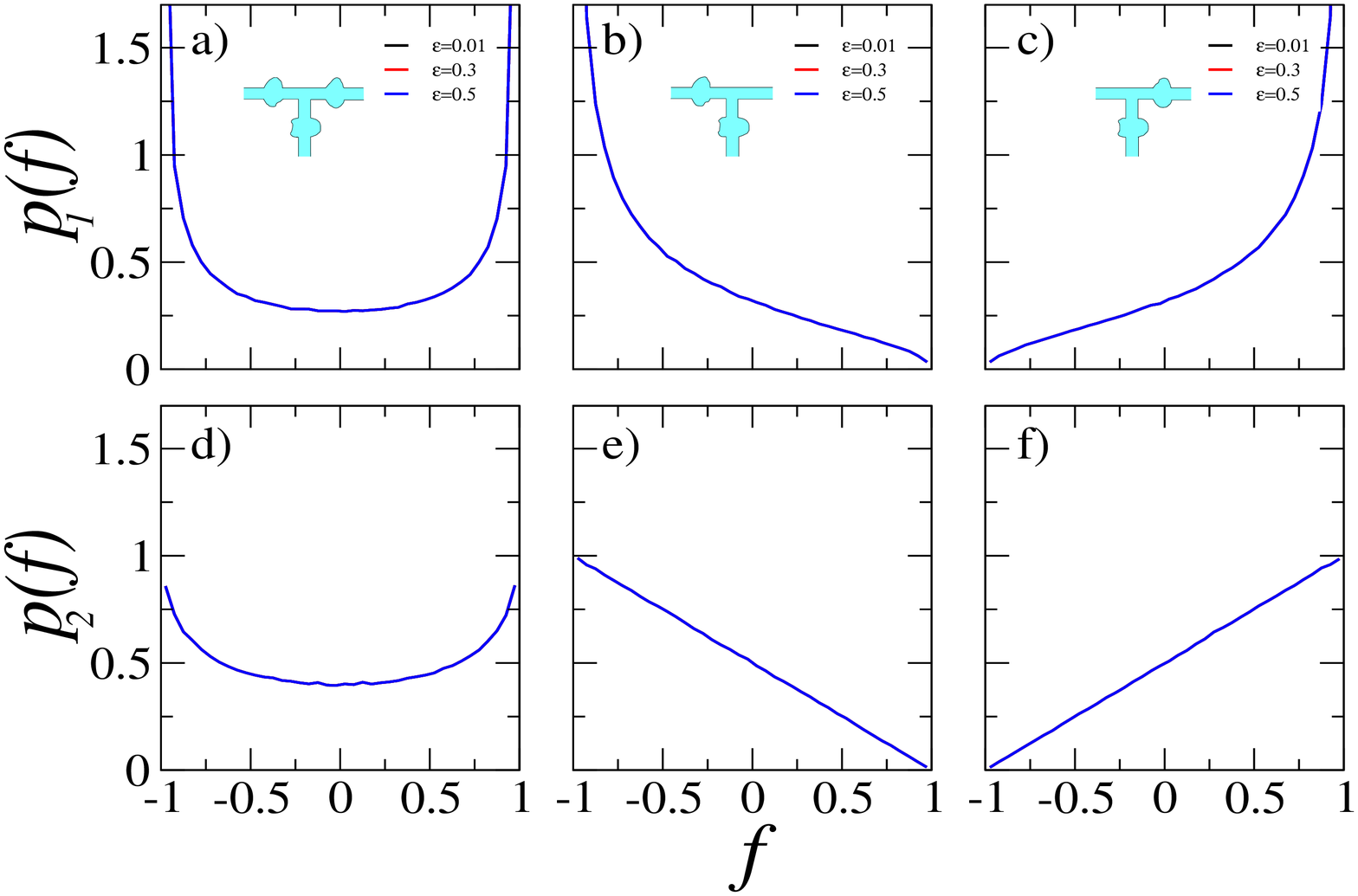}
\caption{Distribution of $f$ for a chaotic three terminal device for different
values of $\varepsilon$ and configurations (insets), (a), (b) and (c) in the
presence, and (d), (e) and (f) in the absence, of time reversal invariance.}
\label{fig:figure2}
\end{figure}

The numerical results for the distribution of $f$, for several values of the
coupling strength $\varepsilon$, are shown in Fig.~\ref{fig:figure2} for
different measurement configurations. Panels (a), (b), and (c) of this figure
are the most general cases of voltage measurements with respect to the position,
where all conductors are chaotic. We can observe a clear dependence on the
position of the probe. In panel (a), the probe is in the middle of the
horizontal wire, and the distribution of $f$ is symmetric around zero, which
means that $\mu_3$ fluctuates symmetrically around the average
$(\mu_1+\mu_2)/2$. However, when the position of the probe changes to one end
of the horizontal wire, the distribution of $f$ is no more symmetric with
respect to zero, as can be seen in panels (b) and (c) of Fig.~\ref{fig:figure2};
in fact, $\mu_3$ tends to be closer to the chemical potential of that
terminal. We also note that the distribution of $f$ is independent of the
coupling parameter. However, when the probe is asymmetrically located in the
horizontal wire, the distribution of $f$ reminds that of the probe in the
midpoint. We can see that our results contrast with the disordered case of
Refs.~\cite{GodoyEPL,Godoy} in both limits of weak and strong disorder.

\subsection{Absence of time reversal invariance}

The scattering matrix $S_j$ for the $\beta=2$ symmetry has the following
parametrization~\cite{MelloLesHouches}, 
\begin{equation}
\label{eq:Ensemble2}
S_{j} = \left(
\begin{array}{cc}
e^{i\phi_{j}} & 0 \\
0 & e^{i\psi_{j}}
\end{array}
\right) \left(
\begin{array}{cc}
-\sqrt{1-\tau_{j}} & \sqrt{\tau_{j}} \\
\sqrt{\tau_{j}} & \sqrt{1-\tau_{j}}
\end{array}
\right) \left(
\begin{array}{cc}
e^{i\phi'_{j}} & 0 \\
0 & e^{i\psi'_{j}}
\end{array}
\right).
\end{equation}
Here, the probability distribution of $S_j$ is given by~\cite{MelloLesHouches}
\begin{equation}
dP_2(S_j) = d\tau_{j} 
\frac{d\phi_{j}}{2\pi} 
\frac{d\psi_{j}}{2\pi} 
\frac{d\phi'_{j}}{2\pi} 
\frac{d\psi'_{j}}{2\pi},
\end{equation}
which defines the \emph{Circular Unitary Ensemble}.

In Fig.~\ref{fig:figure2}, panels (d), (e) and (f), we show the results for the
distribution of $f$ for the same values of $\varepsilon$ and configurations as
in the $\beta=1$ case. The dependence on the intensity of the coupling, as well
as in the position of the probe, is also observed. As in the $\beta=1$ case, the
distribution of $f$ is independent of $\varepsilon$ and has memory with respect
to the measurement in the midpoint of the horizontal wire. What is important to
note here is that the distribution of $f$ is strongly affected by the broken
symmetry of time reversal.


\section{Conclusions}
\label{sec:conclusions}

We studied the voltage drop along a horizontal wire with one open mode,
consisting of chaotic conductors, in a three terminal device. This was done by
using a probe which is chaotic. Our analysis was based on random matrix theory
simulations for the chaotic elements, in the presence and absence of time
reversal invariance. We found a clear dependence of the position of the probe in
the horizontal wire. Also, we found a strong dependence of the time reversal
symmetry.

\acknowledgments

The authors thank the organizers of the V Leopoldo Garc\'ia-Col\'in Mexican
Meeting for their kind invitation. AMM-A acknowledges financial support from
CONACyT, Mexico. MM-M is a fellow of Sistema Nacional de Investigadores,
Mexico; he also thanks MA Torres-Segura for her encouragement.


\end{document}